# Further Explanation of the Cosmological Constant Problem by Discrete Space-time through Modified Holographic Principle

H.M.Mok

Radiation Health Unit,
Hong Kong SAR Govt.
3/F., Saiwanho Health Centre,
28 Tai Hong Street,
Sai Wan Ho, Hong Kong, CHINA.
e-mail: a8617104@graduate.hku.hk

**Abstract**

The author proposed in his previous papers in 2001 that the problem of the cosmological constant could be resolved and its calculated value agrees excellently with the observations by assuming that the space-time itself, as the phase of Higgs condensate, is discrete in nature. Further to such idea, a different approach involving the concept of vacuum entropy that is constrained by the modified Holographic Principle is being used in the calculation of the cosmological constant value at three different phase transition times in the early universe. It is interesting and quite unexpected that it arrives at the same conclusion as the previous approach. It therefore reveals that something fundamental is present in the theory. The problem on the preservation of Lorentz invariance in discrete space-time is also discussed.

# Content

## Introduction

We are indeed in the golden era of cosmology [1,2]. In the past decade, tremendous advances in astrophysical observations provided significant evidence for testing the theoretical cosmological model. The revolutionary discovery of the accelerating universe based upon the measurements of SNIa in 1998 [3], not just resolve the age crisis of the universe but, also opened a new page on cosmology. By combining the data from other cosmic probes, for instance, the CMB anisotropy measurement by Boomerang, Maxima, DASI and WMAP and the galaxy red-shift surveys on large scale structure by 2dF and SDSS, the cosmological parameters are properly constrained and converged to a consistent standard cosmological model, or so called the concordance model [4]. It describes a universe of flat geometry and experienced inflation in its early time after birth. It comprises of matter and dark energy of about 0.3, including about 5% baryonic matter and 25% dark matter, and 0.7 of critical mass density respectively in the present epoch. The age of the universe has been also pinned down to 13.7 Gyr [5]. Cosmology is now rapidly evolving into a rigorous, quantitative branch of astrophysics of high precision. It is expected that the cosmological model will be further confirmed and refined by the future satellite mission, such as PLANCK, SNAP etc, and also through ground based observations. The quest on the nature of both the dark energy and dark matter are high on the list of, not just in astrophysics research, but also the physics community.

Although significant progress has been made for the cosmological model, the fundamental cause of the cosmic acceleration is still an open question. The properties of such acceleration could be summarized in a general term named dark energy. It is a special energy form characterized by an equation of state with a negative pressure to energy density ratio $w$. Such ratio might be time dependent in most of the quintessence models and it behaves as the cosmological constant if $w = -1$. So far, its value, as constrained by the combined data of the most recent cosmological observation, is consistent with the scenario of cosmological constant [6] although other quintessence models may not be completely ruled out. The determination of $w$ to high precision <5% and its time dependence are the crucial observational challenges in next decade. If it is concluded that $w = -1$ and behaves as a constant, we may be confronted by the long standing mystery of the cosmological constant problem as the major difficulty.

The author proposed in his previous papers in 2001 [7] that the problem of the cosmological constant could be resolved and its calculated value agrees excellently with the observations by assuming that the space-time itself, as the phase of Higgs condensate, is discrete in nature. The Higgs field described by the electroweak theory could just be the field of the space-time condensate and it agrees with the property that the Higgs field pervades uniformly in space. If space-time is conceived as the

Higgs condensate, the nature of negative pressure with positive vacuum energy could be easily understood and explained. The inconsistency on the scale of the microscopic vacuum energy and its cosmological counterpart could also be reconciled. Particles traveling in free space are dispersed by the space-time condensate (i.e. the Higgs condensate in electroweak theory) as the usual relativistic energy-momentum relation through the mass generation mechanism of the electroweak theory. The phase transitions of the universe were corresponding to the change of the binding energy of the space-time condensate. The cosmological constant values of the three phase transition times in the early universe were found and the inflation may be also attributed to the cosmological constant effect through phase transition. If there will be no phase transition further, our universe may accelerate continuously and eventually ripped apart [8].

Further to such idea, a different approach involving the concept of vacuum entropy constrained by the modified Holographic Principle is being used in the calculation of the cosmological constant value at the three different phase transition times in the early universe. It is interesting and also quite unexpected that it arrives at the same conclusion as the previous approach. It therefore reveals that something fundamental is present in the theory.

**Vacuum Entropy**

If the space-time could be conceived as condensate and discrete in nature, entropy might be associated to it as our usual understanding in condensed matter. Such vacuum entropy, as we may call it, might be closely related to the basic building blocks, or say the constituents, of the space-time condensate which could be treated as information bits. We may not further restrict ourselves to specific model of space-time discreteness, such as spin network, random lattice, etc. as it is irrelevant to our discussions. Analogous to the information entropy, the vacant constituent which acts as the discrete vacuum, could be defined as the "zero" state while the one occupied by matter could be defined as the "one" state irrespective of the spin or particle species, that may contribute more degree of freedom, for simplicity. The physical world could then be conceived as an immense collection of information with energy and matter as the incidentals and all physical phenomenons become a kind of information processes. The maximum vacuum entropy that can be contained in certain region of space is governed by the Holographic Principle (HP). As originally postulated by 't Hooft and Susskind [9,10], the HP provides a physical cut-off mechanism for the energy of the quantum field other than the Planck energy. It also requires that the entropy of a spatial region reside not in the interior but on the surface of the region as storing image information by hologram in optics. Furthermore, the number of degree of freedom allowed per unit surface area of such region is no greater than one per Planck area.

The energy required in confining matter in the space-time vacancy, that is equivalent to the information storage in our interpretation, is governed by the quantum mechanical principle. The relationship between the required energy $E$ and the size of the constituent $x$ is $Ex \sim 1$ ($\hbar = c = 1$ is assumed throughout the paper). As the size of the space-time vacancy is assumed to be in Planck scale in our hypothesis, the energy required to confine a particle in a single vacancy should also be in Planck scale. Such particle will itself form a microscopic Planck black hole and, according to the HP, it will contribute one unit of entropy since the maximum entropy contained is equal to the surface area in Planck units. The maximum number of such microscopic black holes that can be filled in a region of space may therefore determine the vacuum entropy. One may then obviously conclude that the vacuum entropy contained in a spatial region is equal to the corresponding number of constituents in general. However, such argument may not be consistent with our hypothesis that the space-time behaves as the Higgs condensate which is the mass generation medium in electroweak theory. It is because if the particle mass is generated through the space-time condensate, the maximum mass energy that can be accommodated into the space-time vacancy should be smaller than the binding energy of the condensate. This can be referred to the coupling constants in the electroweak theory. It can also be understood by the fundamental reason that if we imagine an extremely strong coupling between the particle and the constituent, they will firmly attach with each other, therefore, the maximum mass scale of the particle should be similar to the binding energy scale of the space-time condensate. In our theory, the accommodation of particle mass larger than the electroweak energy scale of the space-time condensate is possible but it may require the condensate to change its phase to higher energy scale, for instance, from the electroweak scale to the GUT scale or even the Planck scale.

If the space-time vacancy occupied by matter energy is smaller than the Planck scale, it may not contain a whole bit of information because it is possible for the particle to exist in some other vacancies due to the quantum uncertainty. The occupancy of each vacancy might be described as the superposition of the "0" and "1" states. In this case, the quantity of vacuum entropy could not be found simply by counting the number of constituents but might be determined by other principle. The consideration of holographic bound provides clues to us on this problem. Although it is not appropriate to assume Planck scale occupancy of the space-time vacancies as mentioned, we may still start from it and then convert the problem to lower energy cases. If we consider a spatial region fully occupied by Planck scale particles of energy density $\rho$ and, at certain radius $R$, the matter energy contained in that region is large enough to meet the black hole condition as

$$G\rho R^2 \sim 1 \qquad (1),$$

then for the vacancies filled by energy scale less than the Planck scale, the corresponding black hole horizon radius will need to be increased in order to contain more bounded mass to meet the black hole condition. Consequently, the larger $R$ value will increase the number of space-time constituent bounded by that radius. Although the number of vacancy increases as $R^3$, the HP constrains the maximum entropy to increase only as $R^2$. The actual number of state allowed per individual space-time vacancy is then varies as $1/R$. By using the black hole condition in equation (1), we have

$$R \sim 1/\sqrt{G\rho} \qquad (2)$$

If we denote the binding energy scale of the condensate by $m$, for Planck scale occupancy in the electroweak condensate, we have $\rho = m^3 M$. On the other hand, for electroweak scale occupancy in electroweak condensate, we have $\rho = m^4$. We may then find the ratio of the number of state allowed per individual space-time vacancy in the case with Planck energy occupancy and the case with electroweak energy occupancy as

$$R_{Planck} / R_{electroweak} = \sqrt{G\rho_{electroweak}} / \sqrt{G\rho_{Planck}} = \sqrt{m^4 / m^3 M} = \sqrt{m/M} \quad (3)$$

It could be known as the "entropy contribution factor" for the constituent on the vacuum entropy. The vacuum entropy can be then found by assuming Planck occupancy and convert to the appropriate energy scale by multiplying this factor on the number of constituents as

$$S_v = \rho_n cV \quad (4)$$

where $S_v$ is the vacuum entropy; $\rho_n$ is the number density of the constituents; $c$ is the entropy contribution factor and $V$ is the volume of the concerned region. That means apart from the case of Planck scale occupancy, the information contain in a region should be less than the interior number of constituents but weighed by the entropy contribution factor.

**Cosmological Constant**

After establishing the concept of vacuum entropy and its contribution factor, let us apply the theory to the entropy of the entire universe. The extension of HP to the universe as a whole was first discussed by Fischler and Susskind (FS) [11] and the subsequent modifications [12]. In the FS version of the cosmic holographic principle (CHP), the particle horizon is assumed to be the boundary for applying the holographic bound. It is reasonable to employ this interpretation as no information outside the horizon could be communicated with the inside or vice versa. Using the above results, vacuum entropy could be independent of whether matter exists or not but associated to the number of constituents in a region and the entropy contribution factor. The vacuum entropy of the whole universe can be constrained by the HP on its horizon area in Planck units and the maximum allowed entropy could be written as

$$A = 4\pi R^2 M^2 = S_v = \rho_n cV = (4\pi/3)(mR)^3 \sqrt{m/M} \qquad (5)$$

where $A$ is the surface area of the horizon in Planck units. The $R$ given by the above equation is then the horizon size that the entropy contained in the bulk approach its maximum value as given by the HP. $R$ can be found as

$$3R^{-1} = \sqrt{m^7/M^5} \qquad (6)$$

However, the expansion of the universe continuously increase the horizon size as

$$R(t) = a(t) \int dt'/a(t') \qquad (7)$$

where $a(t)$ is the scale factor and $t$ is the cosmic time. The surface area to bulk volume ratio is therefore varies as $1/R$. The holographic bound will then be violated when the entropy contained in the bulk exceeds the maximum allowed by the HP at the horizon size given by equation (5). In a matter dominant universe, one can find that $R = 3t$. However, interestingly, if the cosmological constant exists, its contribution will gradually dominate the energy density and cause the acceleration of the universe. In a universe dominated by the cosmological constant (i.e. $\Omega_m = 0$), its horizon size is varied as

$$R(t) = \left(1 - \exp(-Ht)\right)/H \qquad (8)$$

If t tends to infinity, the horizon size approaches its limiting value as $R = 1/H = \lambda^{-1/2}$ and the surface to volume ratio become constant. Therefore, the HP will be protected [13] and the situation can be rescued. For estimating the time required to change from a matter dominated universe to a $\lambda$ dominated one, we know from the FRW equation that in a matter dominated universe

$$t = 2a_c \mu^{3/2}/3 \qquad (9)$$

where $\mu = a/a_c$ and $a$ is the scale factor and

$$a_c = 8\pi G \rho_0 a_0^3 / 3 = 8\pi G \rho a^3 / 3 = H^2 \Omega_m a^3 \qquad (10)$$

When the energy density contributed by the cosmological constant is comparable to the matter density, we can take $\Omega_m = \Omega_\lambda$ and write $a_c = \lambda a^3 / 3$. So that,

$$t \sim \lambda^{-1/2} \qquad (11)$$

The horizon size at that time was then

$$R \sim \lambda^{-1/2} \qquad (12)$$

We therefore know that the horizon size at the time of domination by the cosmological constant is of similar order with its limiting value for the protection of HP by the acceleration effect.

By substituting the $R$ value into the equation (6), we have

$$\lambda \sim m^7 / M^5 \qquad (13)$$

As mentioned in the previous paper, if there were three transition energy scale, those are the Planck scale, the GUT scale and the electroweak scale, associated to the phase changes of the space-time condensate in the early universe, of the universe, we may find the corresponding horizon size which attained the maximum entropy of the universe by the above equation (6). The corresponding cosmic time could also be found as $10^{-42}$s, $10^{-27}$s and 8 billion years of the universe respectively and the universe would then dominate by the cosmological constant as given by equation (13) for the corresponding phase of the condensate. One may find that the cosmological constant values for the phases of Planck and GUT are enormously large that could be the origin of inflation in the early universe and such fast expansion then triggers the phase transition of the condensate. The last one is excellently agreed with the most recent observation that the cosmological constant dominated the energy density of the universe at 5-6 billion years ago [14]. The above calculation shows that the cosmological constant and the calculated values at these phases of the universe are well agreed with those found by the approach as mentioned in my previous paper. Such agreement on the results is quite unexpected because they are based on different theoretical foundation.

Concerns may be raised on the problem of preservation of Lorentz invariance in discrete space-time, however, the quantum mechanical nature of the constituents rescues us on the problem. Let us draw an analogy by the translation symmetry for a particle in free space. If we treat such particle classically, its probability distribution will behave as a point Dirac function and obviously it breaks the translation symmetry of free space. However, if it is a quantum particle in the momentum eigenstate, the translation symmetry is preserved because of the traveling wave behaviour of its wavefunction. Similarly, for the case of Lorentz invariance, we can follow the same idea but include further the concept of path integral. Motion of a classical particle in discrete space-time will be represented by a zig-zag path through discrete points in the space-time diagram and obviously it breaks the Lorentz symmetry. However, for a quantum particle moving on the discrete space-time condensate of quantum mechanical nature, we have to sum up the phase factors of all the possible configuration of the space-time discrete points of the condensate and also all the possible path through such discrete points. One can show that the result will be the same as the path integral description of a quantum particle in continuous space-time because the difference is just the approach of summing over the whole space-time of the integral. That means if the lagrangian of both the Higgs condensate and the particle are Lorentz invariance, the final path integral is Lorentz invariance. Combining this view with the electroweak theory, the usual relativistic energy-momentum equation is just the dispersion relation for the particle traveling in the condensate as the masses of particles are generated by it.

**Discussions and Conclusion**

Starting from the discrete space-time as the fundamental hypothesis, we have established the concept of vacuum entropy. By making use of the HP, the discussion of the maximum amount of energy contained in space-time through microscopic Planck black hole consideration generates the concept of entropy contribution factor. This factor helps us in determining the quantity of vacuum entropy from the number of space-time constituents by converting the Planck scale occupancy of the space-time vacancy to the appropriate energy scale. From the constraint on the horizon size by HP for the entire universe, we can find out the cosmological constant through the FRW equation. It is interesting to note that, although the cosmological constant as a free parameter in the Einstein equation is geometrical in nature, we show from the above that the introduction of the vacuum entropy and the HP, which both describe the thermodynamic behaviour of the vacuum, leads to its existence and also determines its value.

On comparing the approach in this paper and the previous one, firstly, they are both originated from the microscopic view of the space-time. The previous approach was started from a Planck scale constituent as the fundamental unit (i.e. no matter and energy at all) and apply a weighing factor to project and average out the electroweak binding energy scale to this constituent to get the macroscopic vacuum energy density. While it treated the microscopic vacuum energy as the internal energy of the discrete space-time, the approach describes here is on the microscopic vacuum entropy. Both the internal energy and entropy are the crucial concepts in describing the thermodynamical properties of a system. Our knowledge in statistical mechanics tells us that it is the microscopic picture (i.e. the discreteness of matter) that makes the macroscopic $2^{nd}$ law of thermodynamics understandable. The situation in our theories is similar. That is the concept of space-time discreteness provides the physical meaning to the vacuum entropy and internal energy and makes them understandable.

Although the fundamental principle employed is different, the same conclusion is arrived. It further shows that the introduction of entropy to the FRW equation and the consideration of HP require the existence of the cosmological constant and its domination in the expansion of the universe. Due to the above reasons, this approach provides a strong support for the notion of discreteness of the space-time and its Higgs condensate behaviour. It may reveal that we are on the right track.

**Reference**

[1] Turner M.S., [astro-ph/0108103] and Int.J.Mod.Phys. A17 (2002) 3446-3458 [astro-ph/0202007]

[2] Carroll S.M. [astro-ph/0107571]

[3] Perlmutter S. et al., Astrophys. J. **517**, 565 (1999)

[4] Lahav O, The proceedings of the XXXVII Rencontres De Moriond "The Cosmological Model" (2002) [astro-ph/0208297] & [astro-ph/0205382]

[5] Nolta M.R., Wright E.L., Page L., et al [astro-ph/0302209]

[6] Bean R., Nucl. Phys. Proc. Suppl. 110 (2002) [astro-ph/0201127] & Phys. Rev. D65 (2002) 041302[astro-ph/0110472]

[7] Mok H.M., Proceedings of the 18[th] Institut d'Astrophysique de Paris Colloquium "On the Nature of Dark Energy" (2002) [astro-ph/0105513]

[8] Caldwell R., et al [astro-ph/0302506]

[9] t'Hooft G., "Dimensional Reduction in Quantum Gravity" edited by A.Ali, J.Ellis and S. Randjbar-Daemi, Salamfestschrift: A Collection of Talks (World Scientific, Singapore, 1993) [gr-qc/9310026]

[10] Susskind L., J. Math. Phys. **36** (1995) 6377 [hep-th/9409089]

[11] Fischlet W. & Susskind L., [hep-th/9806039]; R.Bouusso, JHEP 9907 (1999) 004 [hep-th/9905177]

[12] Carneiro S., [gr-qc/0206064]

[13] Cardenas V.H., [gr-qc/0205070]

[14] Reiss A.G., et al., Astrophys. J. (in press) [astro-ph/0402512] & [astro-ph/0104455]